\documentstyle[epsf]{elsart}
\begin{document}
\begin{frontmatter}
\begin{flushright}
  {\bf TTP96-55\footnote{The complete paper, including figures, is also
      available via anonymous ftp at
      ftp://ttpux2.physik.uni-karlsruhe.de/, or via www at\\
      http://www-ttp.physik.uni-karlsruhe.de/cgi-bin/preprints/}\\
    MPI/PhT/96-122\\
    hep-ph/9611354\\
    November 1996}
\end{flushright}
\title{AUTOMATIC COMPUTATION OF THREE-LOOP TWO-POINT FUNCTIONS IN
    LARGE MOMENTUM EXPANSION\thanksref{tfoot}}
\thanks[tfoot]{Extended version of the contribution to the proceedings of the
      AIHENP96 workshop in Lausanne, September 1996.}
\author[MPI,INR]{K.G. Chetyrkin},
\author[TTP]{R. Harlander},
\author[TTP]{J.H. K\"uhn} and
\author[MPI]{M. Steinhauser}
\address[MPI]{Max-Planck-Institut f\"ur Physik,
  Werner-Heisenberg-Institut, 80805 Munich, Germany}
\address[INR]{Institute for Nuclear Research, Russian Academy of
  Sciences, Moscow 117312, Russia}
\address[TTP]{Institut f\"ur Theoretische
  Teilchenphysik, Universit\"at Karlsruhe, 76128~Karlsruhe, Germany}
\begin{abstract}
We  discuss  the  calculation of two-point three-loop functions with an
arbitrary number of massive propagators  and one large external momentum. The
relevant subdiagrams are generated automatically.  The resulting
massless two-point integrals and massive tadpoles are transformed
on-line to FORM-expressions ready to be used by existing FORM packages
which calculate them analytically.  As an example we compute  the
quartic mass corrections to the photon polarization function.
\end{abstract}

\end{frontmatter}

{\bf1 Introduction.}
In general it is not possible to compute analytically a three-loop
propagator diagram with non-vanishing external momentum and internal
particles with different masses. However, often a hierarchy between
the mass scales exists.  A typical example is the case when the
magnitude of the external momentum, $q$, is much larger than all the
masses appearing in the diagram.  For this kinematical situation there
exists a systematic procedure, the Large Momentum Expansion (LME),
which works on the diagrammatic level and delivers an expansion in
$m^2/q^2$ for a given dimensionally regularized diagram \cite{Smi95}.
Possible physical applications of LME at three-loop level include e.g.
the production of $t\bar{t}$ pairs at $e^+e^-$ colliders and the quark
mass effects in the decay rates of Z and Higgs bosons .

In this work we discuss a computer-algebra implementation of the LME
at three-loop level. Note that the one- and two-loop cases are
technically much simpler.  The number of various prototype diagrams to
be considered is rather small and all necessary general diagrammatic
manipulations (except for the final calculation of the very diagrams
produced by LME) are quite possible to do by hand as it was demonstrated in
\cite{DavSmiTau93}.  In contrast to the two-loop case an application
of LME to even  a single three-loop diagram leads to 
dozens of subdiagrams and corresponding Feynman integrals. This makes
the use of  computer algebra methods a must. Once the expansion is
constructed it is to be followed by usual UV renormalization. As the
second step is straightforward we shall only deal with bare Feynman
integrals regulated by taking a generic space time dimension $D = 4 -
2\epsilon$.

We start with listing the rules of LME.
Given a specific diagram with a large external momentum one gets an
asymptotic expansion for the momentum going to infinity by applying the
following prescription:
\begin{enumerate}
\item\label{presc1}
  Generate all subdiagrams of  the initial graph  such that
  \begin{enumerate}
  \item they contain both  vertices through which  the  large 
    momentum goes into and out from the initial graph
  \item they become one-particle-irreducible when the two vertices are
    identified. 
  \end{enumerate}
\item\label{presc2}
  Taylor-expand the integrands of these subdiagrams in all small
  masses and external momenta generated by removing lines from the
  initial diagram. 
\item\label{presc3}
  In the initial diagram, shrink the subdiagram to a point and
  insert the result obtained from the expansion in (\ref{presc2}).
\item\label{presc4}
  Sum over all terms.
\end{enumerate}
We will call the subgraphs from step~(\ref{presc1}) {\it hard subgraphs}
or simply {\it subgraphs}, while the reduced graph, obtained from
step~(\ref{presc3}), will be called the {\it co-subgraph}.  

\begin{figure}
  \begin{center}
  \leavevmode
    \epsfxsize=10.cm
    \epsffile[115 455 450 700]{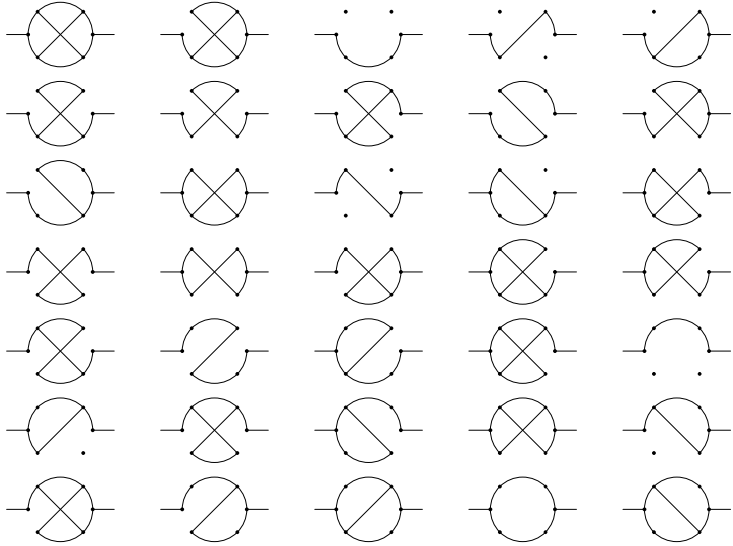}
  \end{center}
  \parbox{13.cm}{
    \caption[]{\label{no_exp_fig} Possible topologies for the subgraphs
      of an NO-type diagram.}
    }
\end{figure}

For the non-planar three-loop topology all hard subgraphs are shown in
Fig.~\ref{no_exp_fig}. 
The corresponding co-subgraphs are obtained from
the full graph by shrinking the displayed lines to a point.
The number of  terms generated  by LME increases
rapidly with the number of loops in the initial diagram. Also, the
relation between the expansion momenta in the subgraph and the loop
momenta of the co-subgraph becomes non-trivial.  Nevertheless, LME
provides well-defined rules and therefore it should be possible to
automate it.

%%%%%%%%%%%%%%%%%%%%%%%%%%%%%%%%%%%%%%%%%%%%%%%%%%%%%%%%%%%%

{\bf2 Implementation.}
We restrict ourselves to three-loop two-point
functions containing an arbitrary combination of massive and massless
lines. 
In this section the emphasis will be put on the 
generation of the terms according to the prescription of
Sect.~1. Once they are available,
existing program packages written in FORM
\cite{form} and based on the method of
integration-by-parts \cite{CheTka81,Bro92} will be used.

The realization of the LME by a computer can be divided into two steps:
(a) generation of the relevant subgraphs and the corresponding co-subgraphs
  including the determination of their topologies, and
(b) distribution of momenta in the subgraph and co-subgraph 
  respecting the relations between them.
As for step (a), one only considers topologies,
disregarding any properties of lines except their relative
positions. Especially one neglects their momentum, mass, particle
type, etc.
For the representation of a diagram we label its vertices by integers
and specify lines by their endpoints. In that way, a topology is
described by the collection of its lines. 

To generate the subdiagrams 
we first note that the full graph is always one of
the hard subgraphs. The remaining ones are obtained by
going through the following steps:
\begin{itemize}
\item Remove any combination of lines from the initial diagram.
\item Remove the emerging isolated dots and binary vertices.
\item Relabel the remaining vertices by 
  $\{1,\ldots,\#{\rm vertices}\}$ and build all permutations.
\item Compare with a table listing the basic one- and two-loop topologies.
\end{itemize}
If a subgraph passes the last step, the
topology of the corresponding co-subgraph is determined in a similar
way.  The result of this procedure is a database containing all relevant
sub- and co-subgraphs, including information about their topology. Note
that no selection criteria bound to line properties have been applied so
far.  Thus, up to this point the procedure is universal and for each
topology this part of the program has to be run only once and for all.

Coming to step (b), we now use the database resulting from (a) and
specialize to a specific diagram, including masses and momenta. 
We bind the direction of the momentum carried by a line to the order of the
labels representing this line. 
Furthermore, for each subgraph
the small external momenta are guided through it in a way that they
touch as few lines as possible.
The final result is the full input for the
FORM-packages, together with all the necessary
administrative files like makefiles etc.

%%%%%%%%%%%%%%%%%%%%%%%%%%%%%%%%%%%%%%%%%%%%%%%%%%%%%%%%%%%%

{\bf3 Application.}
We present the result for the photon
polarization function in QED up to three loops and
mass corrections up to ${\cal O}((m^2/q^2)^2)$. The one- and 
two-loop results are known since a long time in analytical form
\cite{KaeSab55}.
A comparison of these one- and two-loop terms up to 
${\cal O}((m^2/q^2)^6)$ calculated by
our program package with the expansion of the exact results 
was successful.
At three loops the result reads:
\begin{eqnarray}
\Pi(q^2) &=& {1\over 16 \pi^2}\,\Bigg\{
       \frac{20}{9} 
     - \frac{4}{3}\ln\frac{-q^2}{m^2}
+\frac{m^2}{q^2}
       8
+\left(\frac{m^2}{q^2}\right)^2
\left(
       4 
      +8  \ln\frac{-q^2}{m^2}
\right)
\nonumber\\
&&
+\frac{\alpha}{\pi}
\Bigg[
       \frac{5}{6} 
     - 4\zeta(3) 
     - \ln\frac{-q^2}{m^2}
+\frac{m^2}{q^2}
\left(
- 12 \ln\frac{-q^2}{m^2}
\right)
\nonumber\\
&&
+\left(\frac{m^2}{q^2}\right)^2
\left(
       \frac{2}{3} 
     + 16\zeta(3)
     - 10\ln\frac{-q^2}{m^2}
     - 12\ln^2\frac{-q^2}{m^2}
\right)
\Bigg]
\nonumber\\
&&
+\left(\frac{\alpha}{\pi}\right)^2
\Bigg[
- \frac{1703}{432} + 
\zeta(2) \left(8 \ln 2 - \frac{23}{3}\right)
- \frac{173}{72} \zeta(3) 
+ 10 \zeta(5) \nonumber \\
& &
+ \left(\frac{47}{24} - \frac{4}{3}\zeta(3)\right) \ln \frac{-q^2}{m^2} 
- \frac{1}{6}\ln^2\frac{-q^2}{m^2}
+ \frac{m^2}{q^2} \bigg(
33 
+ \zeta(2) (48 \ln 2 \nonumber \\
&&
- 46)
+ \frac{7}{3} \zeta(3)
- \frac{70}{3} \zeta(5)
+ \frac{43}{6} \ln \frac{-q^2}{m^2}
+ 7 \ln^2\frac{-q^2}{m^2} \bigg)\nonumber\\
&&
+ \left(\frac{m^2}{q^2}\right)^2 \bigg(
\frac{1214}{27}
+ \frac{640}{9} \zeta(3)
- 48 \zeta(4)
- 20 \zeta(5)
+ 8 B_4  \nonumber\\
&&
+\left(\frac{1003}{36}
+ \zeta(2) \left(96 \ln 2 - 92 \right) 
- \frac{104}{3} \zeta(3)
\right) \ln\frac{-q^2}{m^2}  \nonumber\\
&&
+ \frac{119}{6} \ln^2\frac{-q^2}{m^2}
+ \frac{32}{3} \ln^3\frac{-q^2}{m^2}
\bigg)
\Bigg]
\Bigg\}
\ldots
\label{eqpi}
\end{eqnarray}
where $B_4\approx-1.762800$ \cite{Bro92}.
The imaginary part of Eq.~(\ref{eqpi}) is in agreement with
\cite{CheKue94} which is an important check for our programs. Terms up
to ${\cal O}((m^2/q^2)^6)$ have been calculated by our program package
and successfully compared with the results obtained by different
methods \cite{CheKueSte96}.

To conclude,
the manual effort for calculating a three-loop propagator-diagram in the Large
Momentum Expansion has been reduced to its minimum. This will find
several applications in physics, e.g. the 
${\cal O}(\alpha_{\rm s}^2 (m^2/q^2)^n)$-terms $(n=3,4,\ldots)$
of the photon polarization function.
The use of these terms as input for the Pad\'e method to get the polarization
function in the whole energy range will certainly contribute to a
further improvement and stabilization of the results of 
\cite{CheKueSte96}. From the technical point of view, the program should
be extendible
also to the hard mass procedure, which will be one of the future projects.

\end{document}